\documentclass[aps,prl,twocolumn,superscriptaddress]{revtex4-2}
\usepackage{graphicx}
\usepackage{amsmath}
\usepackage{amssymb}
\usepackage{colordvi}
\usepackage{mathrsfs}
\usepackage{bm}
\usepackage{verbatim}
\usepackage{dcolumn}
\usepackage{epsfig}
\usepackage{subfigure}
\usepackage[colorlinks,allcolors=blue]{hyperref}
\usepackage{ulem}
\usepackage{makecell}

\begin{document}
\title{Quantum Anomalous Hall Effect in $d^{10}$ Oxide Monolayers}
\author{Zeyu Li}
\thanks{These authors contributed equally to this work.}
\affiliation{International Center for Quantum Design of Functional Materials, and Department of Physics, University of Science and Technology of China, Hefei, Anhui 230026, China}
\affiliation{Hefei National Laboratory, University of Science and Technology of China, Hefei 230088, China}

\author{Xudong Zhu}
\thanks{These authors contributed equally to this work.}
\affiliation{Institute of Artificial Intelligence, Hefei Comprehensive National Science Center, Hefei, Anhui 230088, China.}

\author{Yulei Han}
\email[Correspondence author:~~]{han@fzu.edu.cn}
\affiliation{Department of Physics, Fuzhou University, Fuzhou, Fujian 350108, China}

\author{Zhenhua Qiao}
\email[Correspondence author:~~]{qiao@ustc.edu.cn}
\affiliation{International Center for Quantum Design of Functional Materials, and Department of Physics, University of Science and Technology of China, Hefei, Anhui 230026, China}
\affiliation{Hefei National Laboratory, University of Science and Technology of China, Hefei 230088, China}

\date{\today}
\begin{abstract}
Quantum anomalous Hall effect (QAHE) arises from the interplay between magnetic order and spin-orbit coupling, which opens up a topologically nontrivial band gap to host chiral edge states in the absence of magnetic field. So far, magnetic order of QAHE usually originates from partially filled transition-metal $d$ orbitals or correlation-driven moir\'e bands. Here, we propose an experimentally accessible family of two-dimensional oxides, M$_2$DO$_6$ (M = Zn, Cd; D = Se, Te), that can realize QAHE from the half-filled O-$2p$ orbital induced spontaneous ferromagnetism. In M$_2$DO$_6$ monolayers, spin-polarized Dirac points appear at K/K$^{\prime}$ valleys and along $\Gamma$–K/$\Gamma$–K$^{\prime}$ lines. $C_3$ rotational symmetry then generates eight symmetry-related crossings in the first Brillouin zone. Upon gap opening by spin-orbit coupling, each massive Dirac point contributes half Chern number, resulting in a high-Chern-number QAHE phase with $\mathcal{C}=4$. We establish cation deintercalation as a general strategy to activate O-$2p$ ferromagnetism in oxides. Our finding provides a route to realize QAHE from O-$2p$ ferromagnetism and offers design principles applicable to oxygen-based magnetic topology platforms beyond conventional $d$-electron systems.
\end{abstract}

\maketitle

\textit{Introduction.---} Quantum anomalous Hall effect (QAHE) realizes a Chern insulating state with quantized Hall conductance in the absence of external magnetic field, arising from the interplay between spin–orbit coupling and ferromagnetism~\cite{QHE,QAHE-Haldane}. Owing to its dissipationless chiral edge transport, QAHE has attracted continuous interest for both fundamental studies of topological quantum effects and potential applications in low-power electronics~\cite{QAHEreview,half-q}. Despite extensive theoretical proposals, its experimental realization remains limited by the stringent requirement of simultaneously achieving nontrivial band topology and robust magnetism. Since 2004, a broad range of material platforms have been explored, including magnetically doped topological insulators~\cite{Cr-Bi2Se3,Exp-Cr-Bi2Se3,np-Bi2Se3,2dTI}, magnetized graphene~\cite{Rashba-Graphene,Qiao-antiferromagnet,Qiao-Shi}, intrinsic magnetic topological materials~\cite{MnBi2Te4-1,MnBi2Te4-2,MnBi2Te4-3,Exp-MnBi2Te4-1,Exp-MnBi2Te4-2,Exp-MnBi2Te4-3,Edge states}, and correlated insulators~\cite{moire1,moire2,moire3,rhombo-gra}. The corresponding representative material systems are Cr-doped (Bi, Sb)$_{2}$Te$_{3}$~\cite{Exp-Cr-Bi2Se3}, magnetic graphene with Rashba spin-orbit coupling~\cite{Qiao-Shi}, MnBi$_{2}$Te$_{4}$~\cite{Exp-MnBi2Te4-1},  and MoTe$_{2}$/WSe$_{2}$ heterobilayers~\cite{moire3}, respectively. One breakthrough is that the observation temperature of the QAHE has gradually increased from millikelvin to several kelvin, with advances in experimental technology and the evolution of the material platform. It is still highly demanding and promising to raise the QAHE observation temperature in future. Realizing higher-Chern-number QAH phases is another compelling goal, as they support multiple chiral edge modes~\cite{high1,high2,high3}. Such phases generally require engineering multiple gapped band crossings whose topological contributions yield a higher Chern number~\cite{high4,high5,high6,high7,Zeng-high}, placing stringent constraints on the electronic structure and leaving few realistic material candidates~\cite{high9,high10,high11,Doung-high}.

Among the aforementioned QAHE material platforms, spontaneous magnetic order is typically rooted in unpaired transition-metal $d$ electrons~\cite{OsCl3,V2O3,RuI3,quart,LaCl} or electron-electron interactions~\cite{ee1,ee2,ee3,ee4}. Whether magnetism can be activated through a distinct mechanism remains an open question. Over past two decades, several studies have shown that oxides with $d^0$ electronic configurations can also exhibit robust ferromagnetism~\cite{p-o-1,p-o-2,p-o-3}, as exemplified by HfO$_2$ thin films~\cite{HfO2} and perovskite interfaces~\cite{Per-inter-1,Per-inter-2}. The underlying mechanism originates from uncompensated charges in the O-$2p$ manifold, which creates partially occupied hole states and drives spontaneous spin polarization.

Layered transition-metal oxides offer a fertile platform where lattice geometry, electron correlations, and spin-orbit coupling intertwine to generate emergent quantum phases, attracting growing interest, including quantum materials~\cite{Hex-topo} and energy storage~\cite{Hex-review}. Motivated by the possibility of $p$-orbital magnetism in $d^0$ oxides, we focus on layered oxides Na$_2$M$_2$DO$_6$ (M = Mg, Zn; D = Se, Te), in which alkali-metal ions are sandwiched between transition-metal tellurate slabs with a honeycomb arrangement~\cite{A2M2XO6-1,A2M2XO6-2}. Owing to their high-voltage electrochemistry, and fast cation diffusion, these compounds have been widely explored as promising sodium-ion battery materials~\cite{A2M2XO6-Battery-1,A2M2XO6-Battery-2}. A$_2$M$_2$DO$_6$ consists of two A$^+$ cations and a [M$_2$DO$_6$]$^{2-}$ framework. Removal of A ions could introduce holes into the M$_2$DO$_6$ layer, thereby providing a possible route to partially occupied O-$2p$ states and spontaneous magnetization.

In this Letter, we systematically investigate monolayer oxides M$_2$DO$_6$ (M = Zn, Cd; D = Se, Te) by using first-principles calculations and reveal a rare kind of high-Chern-number QAHE driven by unpaired O-$2p$ states. These monolayers can be obtained by advanced exfoliation methods from experimentally realized layered compounds, such as Na$_2$Zn$_2$TeO$_6$, where sodium ions are intercalated between adjacent M$_2$DO$_6$ layers. Their dynamical and thermal stabilities are confirmed by phonon spectra and molecular dynamics simulations, respectively. We find that they are easy-axis ferromagnetic half-metals with wide spin-polarization windows exceeding 0.5 eV. Near the Fermi level, the spin-polarized bands are dominated by O-$2p$ orbitals. In the first Brillouin zone, eight Dirac points related by $C_3$ rotational symmetry emerge near the Fermi level. When spin-orbit coupling is included, these Dirac points are gapped, giving rise to a topologically nontrivial phase with a high Chern number of $\mathcal{C}$=4. In the end, we demonstrate cation deintercalation as a general strategy for activating O-$2p$ ferromagnetism in oxides.

\begin{figure}
	\centering
	\includegraphics[width=0.45\textwidth]{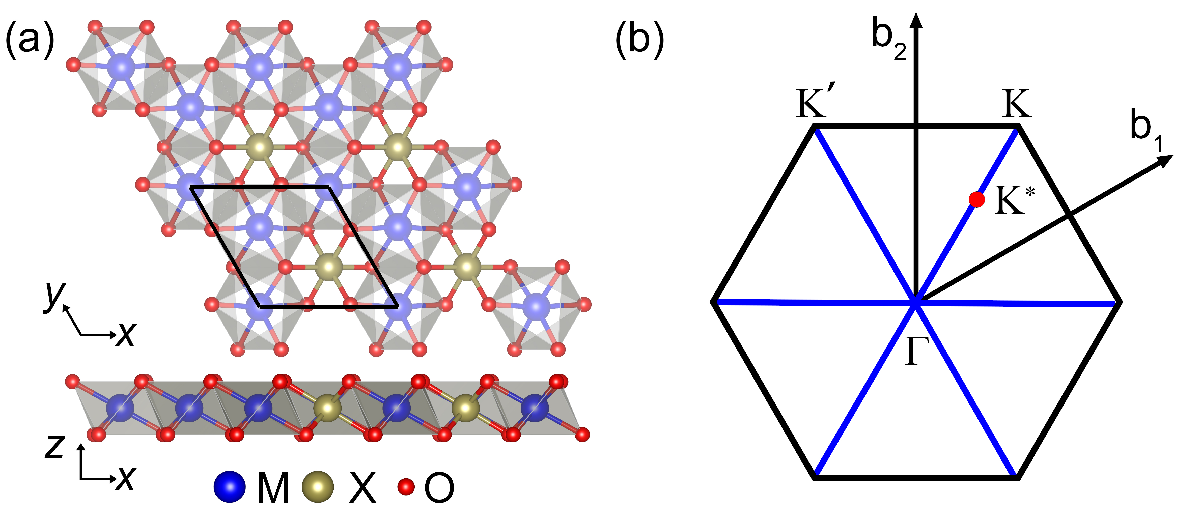}
	\caption{(a) Top and side views of monolayer oxides M$_{2}$DO$_{6}$ (M = Zn, Cd; D = Se, Te). (b) The first Brillouin zone with high symmetry points and high high-symmetry line. The red point K$^\ast$ represents the momentum point where the Dirac dispersion emerge along the K-$\Gamma$ high high-symmetry line.}
	\label{fig1}
\end{figure}

\textit{Structural Properties.---} Inspired by experimentally known honeycomb layered oxides A$_2$M$_2$DO$_6$, we consider monolayer M$_2$DO$_6$ as a derivative obtained by removing the intercalated A ions from the parent compounds, crystallized in $P_3$ space group (No. 143). As shown in Fig.~\ref{fig1}(a), both M and D are octahedrally coordinated by oxygens, forming edge-sharing MO$_6$ and DO$_6$ octahedra that constitute an O-M/D-O trilayer structure. The optimized lattice constants, listed in Table S1  in the Supplemental Materials~\cite{Supplementary}, increase with the atomic numbers of M and D. Phonon calculations based on a 3$\times$3$\times$1 supercell exhibit no imaginary frequencies, confirming the dynamical stability of all monolayer M$_2$DO$_6$ compounds (see Fig.~S1). Molecular dynamics simulations at 300 K show only small total-energy fluctuations, further supporting their room-temperature thermal stability (see Fig.~S2).

\begin{figure}
\centering
\includegraphics[width=0.5\textwidth]{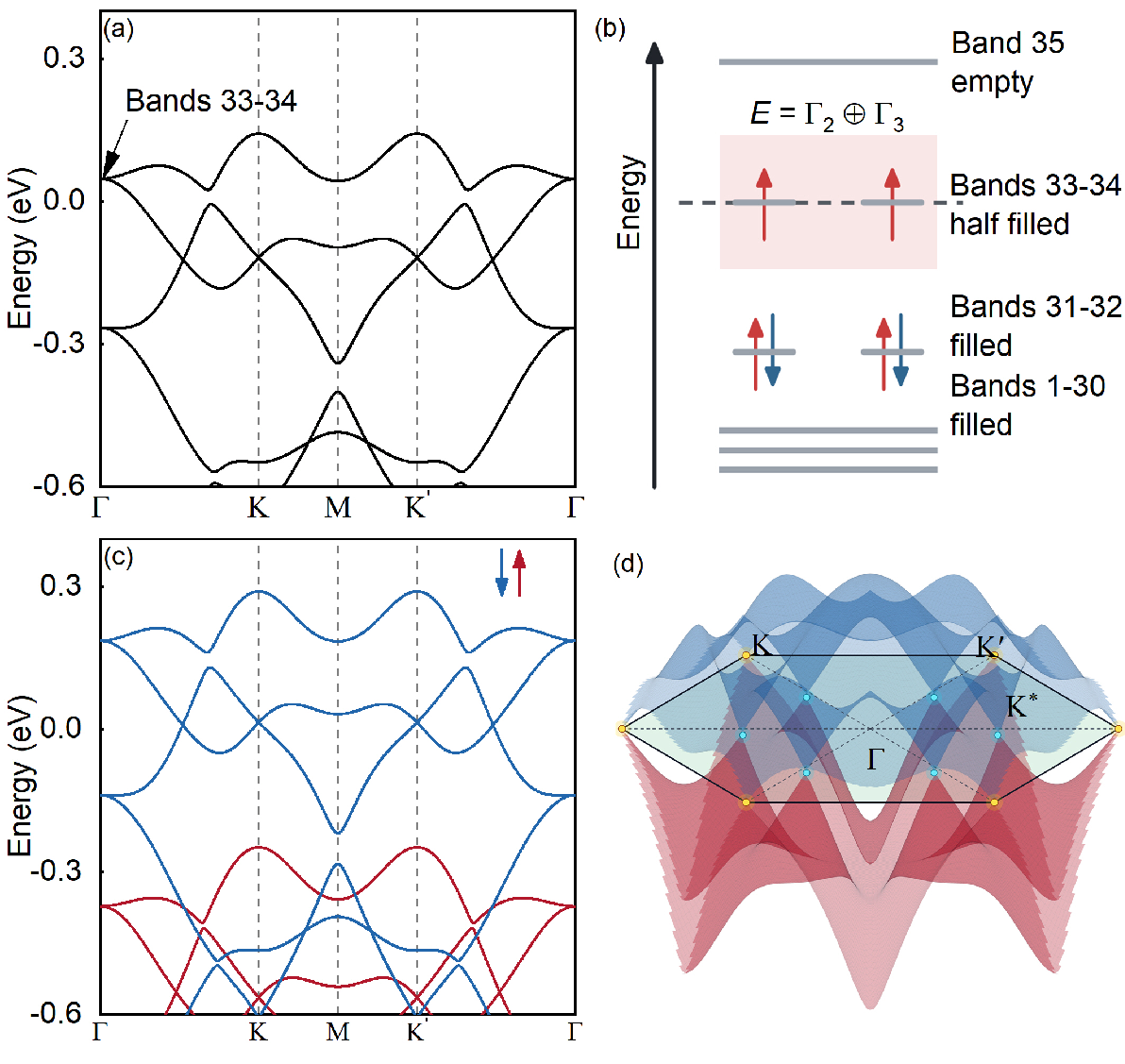}
\caption{(a) Nonmagnetic band structure of monolayer Zn$_2$TeO$_6$. (b) The schematic illustrates the symmetry-constrained filling of nonmagnetic monolayer Zn$_2$TeO$_6$. Bands 1–32 are fully occupied, while bands 33 and 34 form a degenerate doublet at the Fermi level that accommodates the remaining two electrons. According to Hund’s rule, these two electrons occupy different components of the doublet with parallel spins, giving rise to a spin-polarized ferromagnetic ground state without the presence of unpaired $d$ electrons. (c) Spin-polarized band structure of monolayer Zn$_2$TeO$_6$. (d) The corresponding three-dimensional band structure near the Fermi level for (c). Blue and red surfaces denote the two crossing bands, while yellow and cyan markers denote the Dirac points at K/K$^{\prime}$ and along the $\Gamma$–K/$\Gamma$–K$^{\prime}$ directions, respectively.}
\label{fig2}
\end{figure}

\begin{table}[b]
\caption{Irreducible representations and characters of the bands near the Fermi level E$_{f}$ at $\Gamma$ point for the C$_3$ little group; $\Gamma_2\oplus\Gamma_3$ corresponds to the $E$ doublet representation. $\chi(E,C_3,C_3^{-1})$ denotes the characters of the corresponding irreducible representation.}
\centering
\renewcommand\arraystretch{1.2}
\label{tab:gamma_irrep}
\begin{ruledtabular}
\begin{tabular}{cccc}
Band & Degeneracy & \begin{tabular}{c} Irreducible\\representation\end{tabular} & $\chi(E,C_3,C_3^{-1})$ \\\hline
28--29 & 2 & $E$:$\Gamma_2\oplus\Gamma_3$ & (2,-1,-1) \\
30     & 1 & $A$:$\Gamma_1$               & (1,1,1)   \\
31--32 & 2 & $E$:$\Gamma_2\oplus\Gamma_3$ & (2,-1,-1) \\
33--34 & 2 & $E$:$\Gamma_2\oplus\Gamma_3$ & (2,-1,-1) \\
\end{tabular}
\end{ruledtabular}
\end{table}

\textit{Origin of Magnetism.---} A stoichiometric A$_2$M$_2$DO$_6$ crystal can be described by the valence configuration 2$\mathrm{A}^+$ + 2$\mathrm{M}^{2+}$ + $\mathrm{D}^{6+}$ + 6$\mathrm{O}^{2-}$, corresponding to a closed-shell, nonmagnetic insulating state. The A ions donate two electrons per unit cell to the [M$_2$DO$_6$]$^{2-}$ framework, thereby providing charge compensation for the negatively charged oxide layers. Complete A ions removal triggers electronic reconstruction. Because M$^{2+}$ retains its closed-shell $3d^{10}$ configuration and Te is already hexavalent, the resulting oxidation is accommodated predominantly by oxygen. The Bader charge analysis shows that the two holes are mainly originated from O-$2p$ states and are distributed over the oxygen sublattice. M$_2$DO$_6$ therefore hosts two holes in the O-$2p$ manifold per unit cell. Correspondingly, O-$2p$ states dominate the high-energy valence bands near the Fermi level (see Fig. S4). Taking Zn$_2$TeO$_6$ as an example, under $C_{3}$ crystal symmetry, nonmagnetic bands 33 and 34 at $\Gamma$ point belong to the complex-conjugate irreducible representations $\Gamma_2$ and $\Gamma_3$ (see Fig.~\ref{fig2}(a)), respectively, and together constitute a two-dimensional $E$-type O-$2p$ doublet. Therefore, 64 of the 66 valence electrons fully occupy the lowest 32 spin-degenerate bands. The remaining two electrons then occupy the $E$ doublet with parallel spins in accordance with Hund's rule, resulting in an $\uparrow\uparrow$ configuration, as shown in Fig.~\ref{fig2}(b). The two unpaired majority-spin electrons give rise to a net magnetic moment of 2 $\mu_B$ per unit cell. Each of the six O atoms carries about 0.261 $\mu_{B}$, giving a total oxygen contribution of 1.566 $\mu_{B}$, approximately 78 $\%$ of the total moment. In contrast, Zn hosts only a weak induced moment of 0.094 $\mu_{B}$ per atom, whereas Te is weakly antiparallel-polarized with -0.034 $\mu_{B}$. These results identify that it is an O-$2p$-driven ferromagnetic state, rather than a conventional cation-centered local-moment state. The resulting spin polarization separates the majority- and minority-spin channels by approximately 0.56 eV. Among the competing magnetic configurations (see Fig. S3), both N\'eel and zigzag states are relaxed to nonmagnetic solutions with higher total energies, whereas the stripe configuration lies about 70 meV per unit cell above the ferromagnetic state. The out-of-plane magnetic anisotropy energy of 0.18 meV per unit cell further supports the stability of ferromagnetism in monolayer Zn$_2$TeO$_6$. Electronic reconstruction therefore transforms the nonmagnetic insulating parent compound into an O-$2p$-dominated half-metallic ferromagnet. The same behavior occurs in the other three M$_2$DO$_6$ compounds. 

\begin{figure}
\centering
\includegraphics[width=0.5\textwidth]{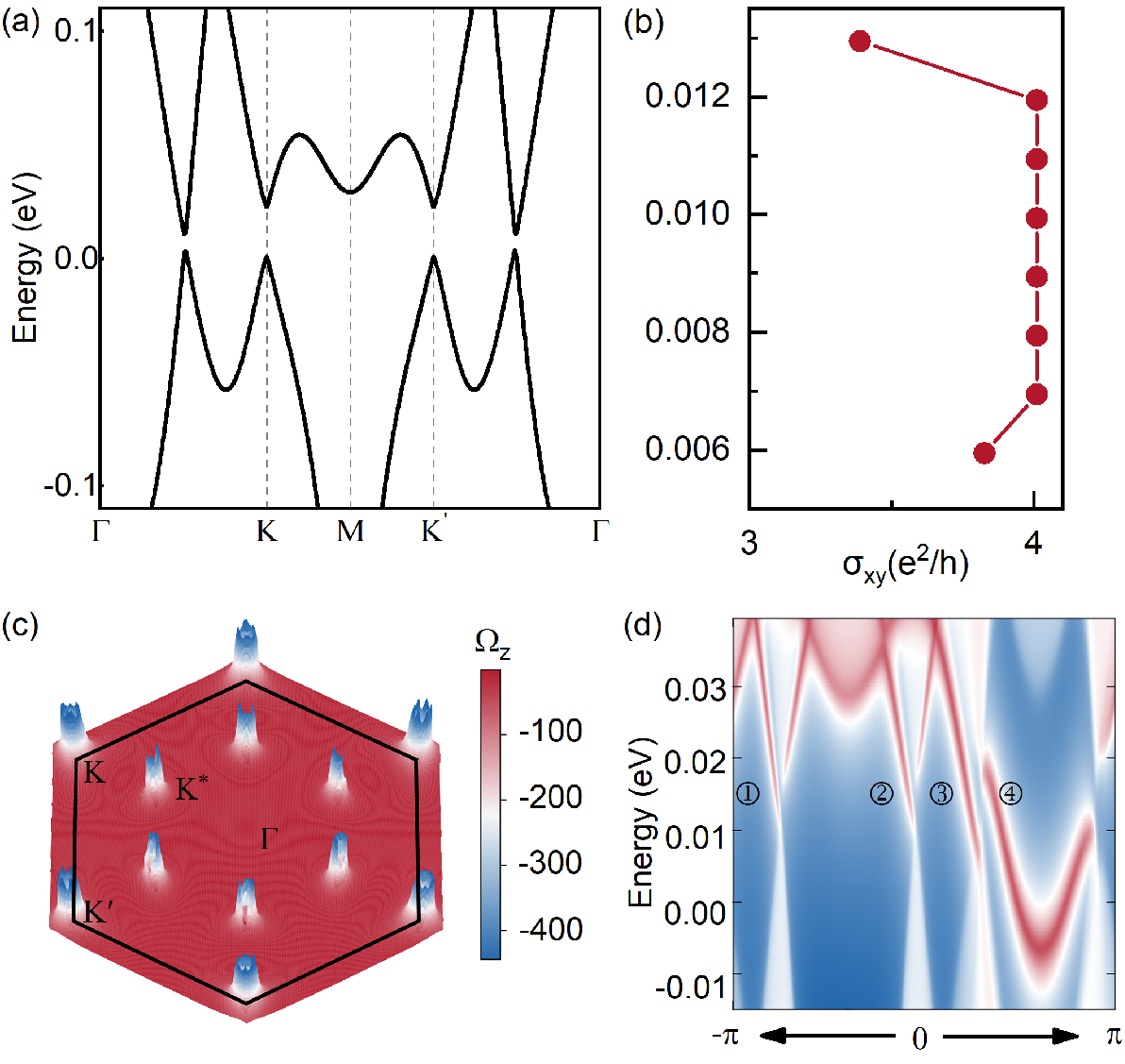}
\caption{(a) Band structure of monolayer Zn$_2$TeO$_6$ with spin-orbit coupling opens a global gap of 7 meV. (b) The calculated anomalous Hall conductance $\sigma_{xy}$ as a function of Fermi energy shows a quantized conductance plateau of 4$e^{2}/h$ in the global gap of monolayer Zn$_{2}$TeO$_{6}$. (c) Corresponding distribution of Berry curvatures mainly concentrates at the gapped Dirac points. (d) Corresponding energy spectra of a semi-infinite ribbon of monolayer Zn$_2$TeO$_6$ with four gapless edge modes.}
\label{fig3}
\end{figure}
\begin{figure}
	\centering
	\includegraphics[width=0.5\textwidth]{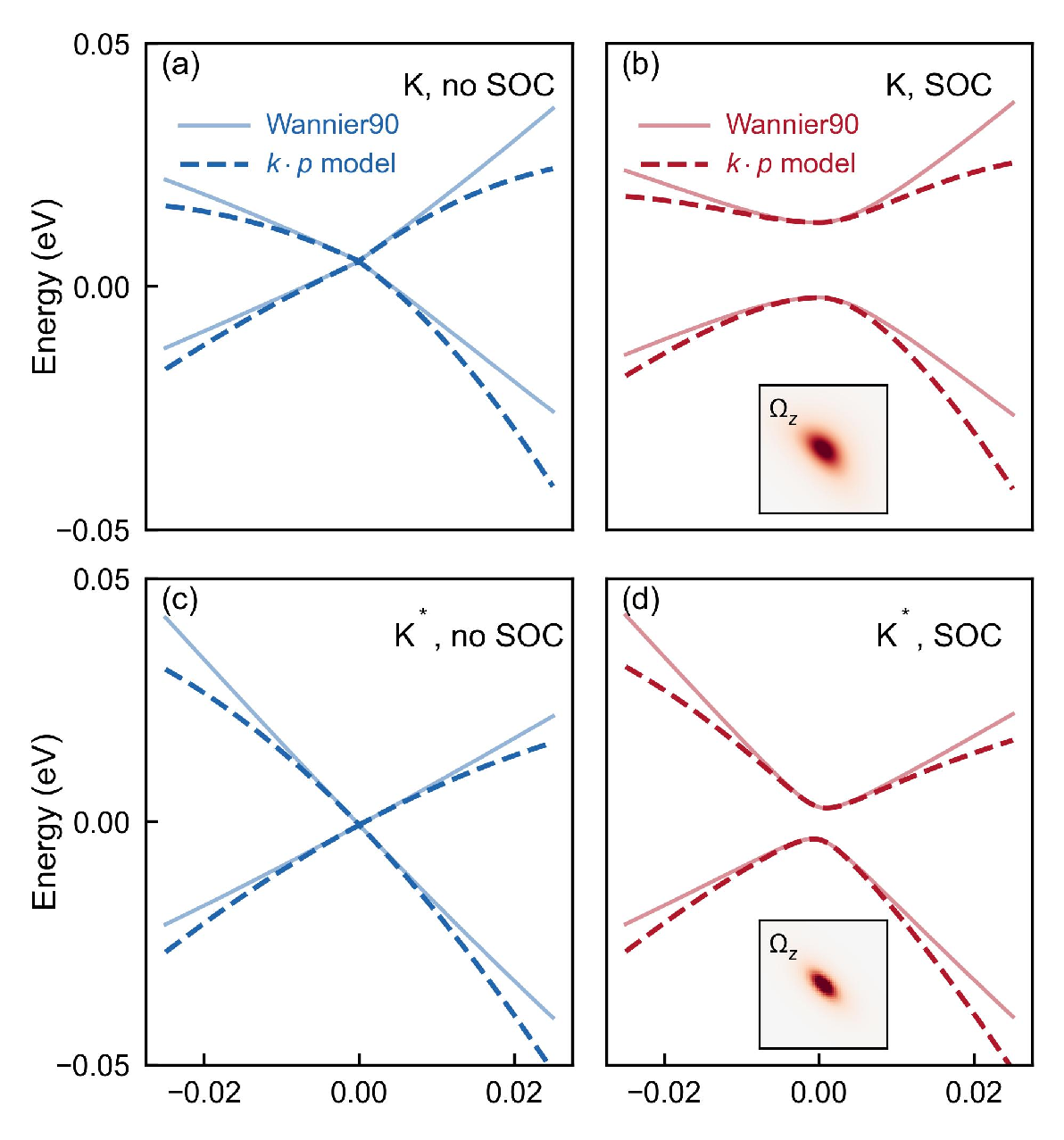}
	\caption{Wannier90 bands and effective two-band k$\cdot$p models near the Dirac crossings at K (a)-(b) and K$^\ast$ (c)-(d). Spin-orbit coupling opens finite mass gaps at both crossings. Insets show the Berry curvature $\Omega_z$ revealing localized topological contributions near the gapped Dirac points.}
	\label{fig4}
\end{figure}

\textit{Band Structures and Topological Properties.---} Taking monolayer Zn$_2$TeO$_6$ as an example, Fig.~\ref{fig2}(c) shows a half-metallic band structure with a spin gap of 0.54 eV. The spin-up channel is gapped near the Fermi level, whereas the spin-down channel remains metallic~\cite{Half-Metallic}, yielding full spin polarization. This suggests potential applications in spin valves and magnetic tunnel junctions~\cite{Spintronics}. Remarkably, spin-polarized Dirac points appear at both K/K$^{\prime}$ valleys and along the $\Gamma$–K/$\Gamma$–K$^{\prime}$ high-symmetry lines (K$^\ast$). Enforced by $C_{3}$ rotational symmetry, these crossings generate eight symmetry-related Dirac points in the first Brillouin zone, as shown in Fig.~\ref{fig2}(d). As shown in Fig. S5, other three M$_2$DO$_6$ compounds also host eight symmetry-related Dirac points near the Fermi level, similar to monolayer Zn$_2$TeO$_6$. In these systems, however, the Dirac points around K$^\ast$ are located below those at K/K$^{\prime}$.

Spin-orbit coupling gaps all spin-polarized Dirac points in monolayer Zn$_2$TeO$_6$, resulting in a global band gap of about 7.0 meV [Fig.~\ref{fig3}(a)]. The Berry curvatures are concentrated near the gapped Dirac points with the same sign[Fig.~\ref{fig3}(c)]. Accordingly, the intrinsic anomalous Hall conductivity exhibits a quantized plateau of 4e$^2$/h within the gap [Fig.~\ref{fig3}(b)], demonstrating monolayer Zn$_2$TeO$_6$ as a high-Chern-number QAHE.

To elucidate the microscopic origin of the high-Chern-number QAHE, we perform an effective two-band k$\cdot$p analysis based on the results of first-principles calculations with and without spin-orbit coupling. Around each crossing point $\mathbf{k}_{c}=(k_{c1},k_{c2})$, we introduce the local momentum $\mathbf{q}=(q_{1},q_{2})$, with $q_{1}=k_{1}-k_{c1}$ and $q_{2}=k_{2}-k_{c2}$ in fractional reciprocal-lattice coordinates. By projecting the full oxygen-$p$-orbital Wannier-based Hamiltonian onto the two crossing bands at $\mathbf{k}_{c}$, we obtain a 2$\times$2 effective model for each Dirac point
\begin{equation}\label{model hamiltonian}
	H_{\mathrm{eff}}(\mathbf{q}) = d_{0}(\mathbf{q})\sigma_{0}+d(\mathbf{q})\cdot\sigma(\mathbf{q}),
\end{equation}
where $\sigma_{0}$ is the two-dimensional identity matrix, $\sigma$ = ($\sigma_{x}$, $\sigma_{y}$, $\sigma_{z}$), and $d$ = ($d_{x}$, $d_{y}$, $d_{z}$). We expand the projected two-band Hamiltonian to linear order in the local momentum, 
\begin{equation}\label{coefficients}
	d_{i}(\mathbf{q})= d_{i}(0) + v_{i1}q_{1} + v_{i2}q_{2} + \mathcal{O}(q^{2}), i=0,x,y,z,
\end{equation}
where higher-order corrections are denoted by $\mathcal{O}(q^{2})$. The coefficient $d_{0}(\mathbf{q})$ determines the center of the two-band spectrum, whereas $d_{x}(\mathbf{q})$ and $d_{y}(\mathbf{q})$ describe the off-diagonal coupling between the two basis states and the anisotropic Dirac dispersion. The term $d_{z}(\mathbf{q})$ acts as a diagonal pseudospin field and provides the Dirac mass at $\mathbf{q}=0$. Figure~\ref{fig4} compares the band structures obtained from the Wannier-based Hamiltonian (solid lines) and the two-band k$\cdot$p model (dashed lines) around the Dirac points at the K/K$^{\prime}$ valleys and K$^\ast$. Figures~\ref{fig4}(a)-(b) show the dispersions around K without and with spin-orbit coupling, respectively, while Figs.~\ref{fig4}(c)-(d) give the corresponding results around K$^\ast$. The effective models accurately reproduce the low-energy dispersions near the Dirac points. Spin-orbit coupling gaps both crossings and generates strongly localized, anisotropic Berry-curvature distributions, as shown in the insets. The local Chern number can be expressed as:
\begin{equation}
	C_{\mathrm{loc}}=\frac{1}{2\pi}\int_{\mathcal{P}}\Omega_{z}(\mathbf{k})\,d^{2}k.
\end{equation}
Each gapped Dirac point carries a local Chern number $C_{\mathrm{loc}}\approx 0.5$. The eight symmetry-related gaps therefore support a high-Chern-number QAHE phase with $\mathcal{C}=4$.

\textit{Effect of Strain.---} In contrast to monolayer Zn$_2$TeO$_6$, the other three compounds, Zn$_2$SeO$_6$, Cd$_2$TeO$_6$, and Cd$_2$SeO$_6$, remain gapless even in the presence of spin-orbit coupling because the Dirac points around K$^\ast$ are lower in energy than those at K/K$^{\prime}$. Given the structural flexibility of two-dimensional systems, strain provides an effective method to tune their electronic, magnetic, and topological properties~\cite{strain1,strain2}. We therefore apply biaxial strain from -5 $\%$ to 5 $\%$ to examine their strain response. Our results show that a small biaxial strain can effectively tune the band structures and bring the eight Dirac points into energetic alignment (see Fig. S5). Taking monolayer Cd$_2$TeO$_6$ as an example, the Dirac points around K$^\ast$ shift upward under 2.5 $\%$ tensile strain and become well aligned with those at the K/K$^{\prime}$ valleys, as shown in Fig. S5(g). After including spin-orbit coupling, these aligned Dirac points open up a topologically nontrivial gap, harboring a high-Chern-number phase with $\mathcal{C}=4$.

\begin{table}
\caption{Cation-deintercalatable oxides parent compounds, corresponding slabs, magnetic ground states, global gaps (meV) , and Chern numbers. }
\centering
\renewcommand\arraystretch{1.5}
\label{slab_summary}
\begin{ruledtabular}
	\begin{tabular}{ccccc}
		\begin{tabular}{c}Parent\\compounds\end{tabular} & Slab & \begin{tabular}{c}Magnetic\\ ground state\end{tabular} & Gap & $C$ \\
		\hline
		\begin{tabular}{c} $\mathrm{Na_3Cd_2SbO_6}$\\ $\mathrm{Li_3Cd_2SbO_6}$ \end{tabular} & $\mathrm{Cd_2SbO_6}$ & FM & 11.41 & 4 \\
		\hline
		\begin{tabular}{c} $\mathrm{Na_3Mg_2SbO_6}$\\ $\mathrm{Li_3Mg_2SbO_6}$ \end{tabular} & $\mathrm{Mg_2SbO_6}$ & FM & 1.48 & 1 \\
		\hline
		$\mathrm{Na_3Cd_2BiO_6}$ & $\mathrm{Cd_2BiO_6}$ & FM & 3.15 & 3 \\
				\hline
		\begin{tabular}{c}$\mathrm{NaInO_2}$\\$\mathrm{LiInO_2}$\end{tabular}& $\mathrm{InO_2}$ & FM & 20.76 & 1
	\end{tabular}
\end{ruledtabular}
\end{table}

\textit{Generalization of O-2p-induced ferromagnetism.---} Finally, we want to propose a general route to O-$2p$ magnetism in cation-deintercalatable oxides. These compounds can be represented as $A_xM_mO_n$, where $A$ is a removable cation, and $M$ is a nonmagnetic closed-shell $d^0$, $d^{10}$, or even main-group ion. Cation deintercalation injects holes into an otherwise nonmagnetic oxide framework, which can be described as 
\begin{equation} 
A_xM_mO_n \xrightarrow{-\,xA^+} M_mO_n + x h^+. 
\end{equation}
Here $A$ denotes the removable cation, $M_mO_n$ is the oxide framework, and $x h^+$ represents the holes injected into the framework by cation deintercalation. When O $2p$ states dominate the valence-band edge, the induced holes can reside on oxygen ligands and spin-polarize the O $2p$ orbitals, creating magnetic moments without partially filled transition-metal $d$ or rare-earth $f$ shells. Coupled through orbital overlap and framework-mediated exchange, these oxygen holes can promote collective magnetism in a nonmagnetic closed-shell lattice. To test the generality of this mechanism, we apply the same computational protocol to more than 20 cation-deintercalatable oxides (See Table S2), including Na$_3$Mg$_2$SbO$_6$, Na$_3$Cd$_2$SbO$_6$, and Na$_2$InO$_3$ as representative examples in which the $M$ ions are $d^0$, $d^{10}$, and main-group ions, respectively. Remarkably, all examined compounds exhibit O-$2p$-driven ferromagnetism upon cation deintercalation. Moreover, monolayers Cd$_2$SbO$_6$, Cd$_2$BiO$_6$, Mg$_2$SbO$_6$, and InO$_3$ listed in Table~\ref{slab_summary}, host nontrivial Chern numbers (See Fig. S6-S10). These results establish cation deintercalation as a general strategy for activating O-$2p$ magnetism and expand the landscape of magnetic oxides beyond the conventional $d$-electron paradigm.

\textit{Summary.---} We propose experimentally accessible two-dimensional oxides M$_2$DO$_6$ (M = Zn, Cd; D = Se, Te) as intrinsic platforms for realizing a high-Chern-number QAHE. Distinct from conventional topological systems driven by transition-metal $d$-orbital magnetism, the nontrivial bands in these compounds arise predominantly from spin-polarized O-$2p$ orbitals, revealing a rare oxygen-orbital-based topological state. In the absence of spin-orbit coupling, two inequivalent spin-polarized Dirac points emerge along the $\Gamma$–K or $\Gamma$-K$^{\prime}$ high-symmetry lines. Related by $C_3$ rotational symmetry, they generate eight Dirac points in the first Brillouin zone. Gapped by spin-orbit coupling, these Dirac points acquire finite masses, each contributing 1/2 to the total Chern number. Consequently, monolayer M$_2$DO$_6$ realizes a high-Chern-number QAHE phase with $C=4$. Moreover, our results show that cation deintercalation can transform nonmagnetic oxides into O-$2p$ ferromagnets, establishing a general route to oxygen-orbital magnetism. Our findings identify $M_2DO_6$ monolayers as a promising platform for high-Chern-number QAHE and point to cation deintercalation as a broadly applicable pathway for activating magnetism through O-$2p$ orbitals.

\textit{Acknowledgements---.} We acknowledge the financial support from the National Key R\&D Program of China (Grant No. 2024YFA1408103), National Natural Science Foundation of China (Grants No. 12474158, No. 12504195, No. 12504195, No. 12234017, and No. 12488101), Innovation Program for Quantum Science and Technology (2021ZD0302800), Fundamental Research Funds for the Central Universities (WK9990250178), Anhui Initiative in Quantum Information Technologies (AHY170000), and China Postdoctoral Science Foundation (Grants No. 2023M733411 and No. 2023TQ0347). We also thank the Supercomputing Center of University of Science and Technology of China for providing the high-performance computing resources.

\end{document}